\documentclass[runningheads]{llncs}
\usepackage[T1]{fontenc}
\usepackage{graphicx} 
\usepackage{amsmath}
\usepackage{amsfonts}
\usepackage{amssymb}
\usepackage{xspace}
\usepackage{amssymb}
\usepackage{xcolor}
\usepackage{colortbl}
\usepackage{multirow}
\usepackage{booktabs} 
\usepackage[misc,geometry]{ifsym}
\definecolor{mygray}{gray}{.9}

\usepackage{color}
\usepackage[pagebackref=true,breaklinks=true,colorlinks,bookmarks=false,citecolor=blue,linkcolor=blue]{hyperref}
\usepackage{dirtytalk}

\newcommand{\eg}{\textit{e.g}}

\begin{document}
\title{3D Mitochondria Instance Segmentation with 
Spatio-Temporal Transformers}

\author{Omkar Thawakar \inst{1} \and
Rao Muhammad Anwer \inst{1,2} \and
Jorma Laaksonen \inst{2} \and \\
Orly Reiner \inst{3} \and
Mubarak Shah \inst{4} \and
Fahad Shahbaz Khan \inst{1}
}
\authorrunning{Thawakar et al.}

\institute{$^1$MBZUAI, UAE, $^2$Aalto University, Finland \\ 
$^3$Weizmann Institute of Science, Israel \\ 
$^4$University of Central Florida, USA \\
\email{omkar.thawakar@mbzuai.ac.ae}}
\maketitle              

\begin{abstract}
Accurate 3D mitochondria instance segmentation in electron microscopy (EM) is a challenging problem and serves as a prerequisite to empirically analyze their distributions and morphology. Most existing approaches employ 3D convolutions to obtain representative features. 
However, these convolution-based approaches struggle to effectively capture long-range dependencies in the volume mitochondria data, due to their limited local receptive field. To address this, we propose a hybrid encoder-decoder framework based on a split spatio-temporal attention module that efficiently computes spatial and temporal self-attentions in parallel, which are later fused through a deformable convolution. Further, we introduce a semantic foreground-background adversarial loss during training that aids in delineating the region of mitochondria instances from the background clutter. Our extensive experiments on three benchmarks, Lucchi, MitoEM-R and MitoEM-H, reveal the benefits of the proposed contributions achieving state-of-the-art results on all three datasets. Our code and models are available at \url{https://github.com/OmkarThawakar/STT-UNET}.
\keywords{Mitochondria segmentation  \and Spatio-temporal Transformer}
\end{abstract}

\section{Introduction}
Mitochondria are membrane-bound organelles that generate the primary energy required to power the cell activities, thereby crucial for metabolism. Mitochondrial dysfunction, which occurs when mitochondria are not functioning properly has been witnessed as a major factor in numerous diseases, including noncommunicable chronic diseases (\eg, cardiovascular and cancer), metabolic (\eg, obesity) and neurodegenerative (\eg, Alzheimer and Parkinson) disorders~\cite{Mcbride2006,Nunnari2012}. Electron microscopy (EM) images are typically utilized to reveal the corresponding 3D geometry and size of mitochondria at a nanometer scale, thereby facilitating basic biological research at finer scales. Therefore, automatic instance segmentation of mitochondria is desired, since manually segmenting from a large amount of data is particularly laborious and demanding. However, automatic 3D mitochondria instance segmentation is a challenging task, since complete shape of mitochondria can be sophisticated and multiple instances can also experience entanglement with each other resulting in unclear boundaries. Here, we look into the problem of accurate 3D mitochondria instance segmentation.

Earlier works on mitochondria segmentation employ standard image processing and machine learning methods ~\cite{reina2011,lucchi2012,lucchi2014}. Recent approaches address~\cite{oztel2017,chenmicaai2022,li2021advanced}  this problem by leveraging either 2D or 3D deep convolutional neural network (CNNs) architectures. These existing CNN-based approaches can be roughly categorized~\cite{mitoEM_Dataset} into bottom-up~\cite{li2021advanced,chenmicaai2022,roneebergerUNet2015,leek2017,chendcan2016} and top-down~\cite{jansu2018}. In case of bottom-up mitochondria instance segmentation approaches, a binary segmentation mask, an affinity map or a binary mask with boundary instances is computed typically using a 3D U-Net~\cite{ozgun2016}, followed by a post-processing step  to distinguish the different instances. On the other hand, top-down methods typically rely on techniques such as Mask R-CNN~\cite{Maskrcnn2017} for segmentation. However, Mask R-CNN based approaches struggle due to undefined bounding-box scale in EM data volume. 

As discussed above, most recent approaches for 3D mitochondria instance segmentation utilize convolution-based designs within the \say{U-shaped} 3D encoder-decoder architecture. In such an architecture,  the encoder aims to generate a low-dimensional representation of the 3D data by gradually performing the downsampling of the extracted features. On the other hand, the decoder performs upsampling 
of these extracted feature representations to the input resolution for segmentation prediction. Although such a CNN-based design has achieved promising segmentation results compared to traditional methods, they struggle to effectively capture long-range dependencies due to their limited local receptive field. Inspired from success in natural language processing~\cite{vaswani2017attention}, recently vision transformers (ViTs) ~\cite{dosovitskiy2021an,liu2021swin,touvron2021training,khan2022transformers,shamshad2022transformers} have been successfully utilized in different computer vision problems due to their capabilities at modelling long-range dependencies and enabling the model to attend to all the elements in the input sequence. The core component in ViTs is the self-attention mechanism that that learns the relationships between sequence elements by performing relevance estimation of one item to other items.  
Inspired by ViTs and based on the observation that attention-based vision transformers architectures are an intuitive design choice for modelling long-range global contextual relationships in volume data, we investigate designing a CNN-transformers based framework for the task of 3D mitochondria instance segmentation. 

\begin{figure}[t!]
\includegraphics[width=\textwidth]{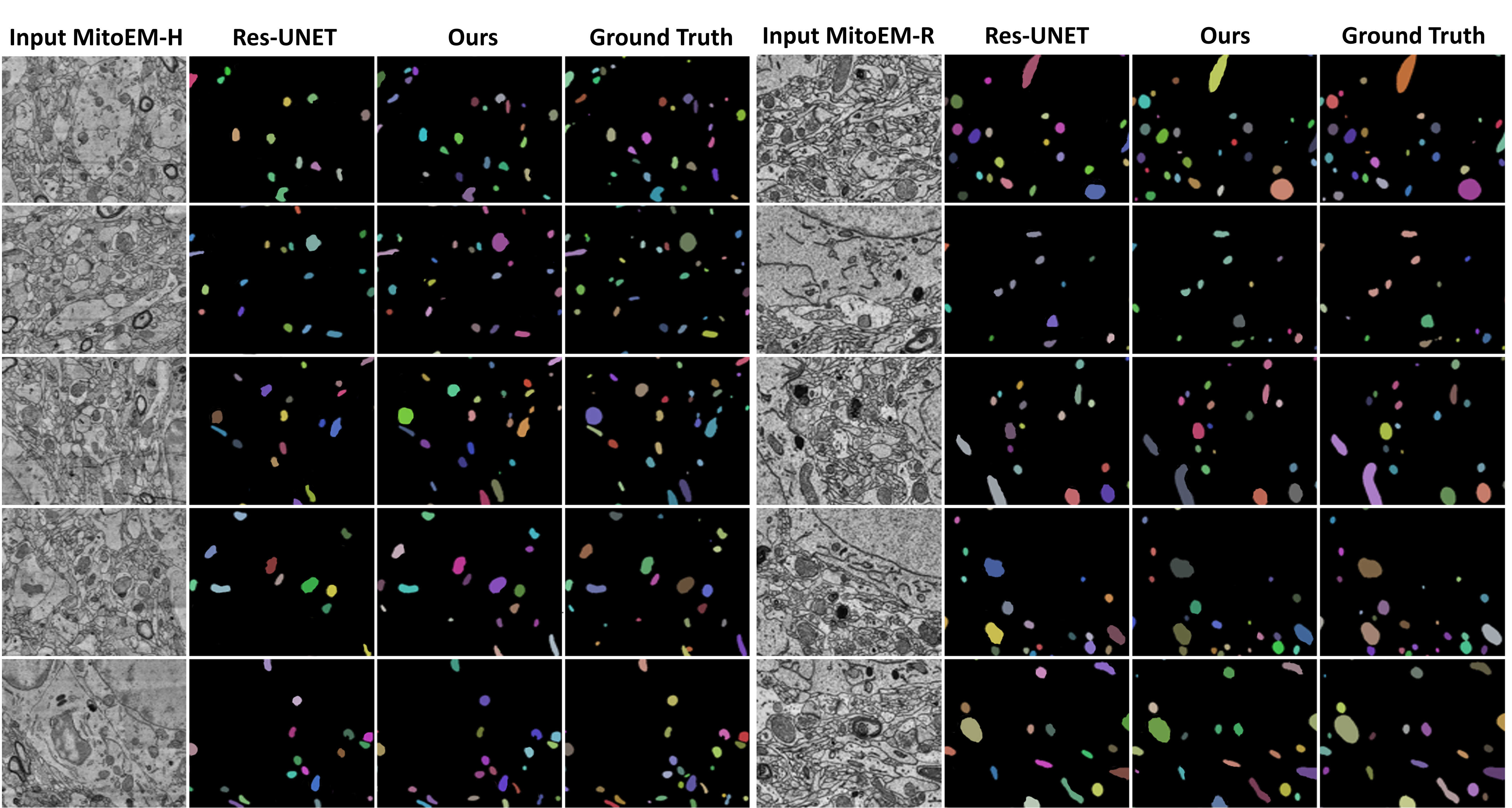} \vspace{-0.65cm}
\caption{Qualitative 3D instance segmentation comparison between the recent Res-UNET~\cite{rsunet} and our proposed STT-UNET approach on the example input regions from MitoEM-H and MitoEM-R validation sets. Here, we present the corresponding segmentation predictions of the baseline and our approach along with the ground truth. Our STT-UNET approach achieves superior segmentation performance by accurately segmenting ~16\% more cell instances in these examples, compared to Res-UNET-R. } \label{intro_figure}  
\vspace{-0.4cm}
\end{figure}

When designing a attention-based framework for 3D mitochondria instance segmentation, a straightforward way is to compute joint spatio-temporal self-attention where all pairwise interactions are modelled between all spatio-temporal tokens. However, such a joint  spatio-temporal attention computation is computation and memory intensive as the number of tokens increases linearly with the number of input slices in the volume. In this work, we look into an alternative way to compute spatio-temporal attention that captures long-range global contextual relationships without significantly increasing the computational complexity. Our contributions are as follows:
 \begin{itemize}
     \item We propose a hybrid CNN-transformers based encoder-decoder framework, named STT-UNET. The focus of our design is the introduction of a split spatio-temporal attention (SST) module that captures long-range dependencies within the cubic volume of human and rat mitochondria samples. The SST module independently computes spatial and temporal self-attentions in parallel, which are then later fused through a deformable convolution.
     \item To accurately delineate the region of mitochondria instances from the cluttered background, we further introduce a semantic foreground-background (FG-BG) adversarial loss during the training that aids in learning improved instance-level features.
     \item We conduct experiments on three commonly used benchmarks: Lucchi~\cite{lucchi2012}, MitoEM-R~\cite{mitoEM_Dataset} and MitoEM-H~\cite{mitoEM_Dataset}. Our STT-UNET achieves state-of-the-art segmentation performance on all three datasets. On Lucchi test set, our STT-UNET outperforms the recent ~\cite{chenmicaai2022} with an absolute gain of 3.0\% in terms of Jaccard-index coefficient. On MitoEM-H val. set, STT-UNET achieves AP-75 score of 0.842 and outperforms the recent 3D Res-UNET~\cite{rsunet} by 3.0\%. Fig.~\ref{intro_figure} shows a qualitative comparison between our STT-UNET and 3D Res-UNET~\cite{rsunet} on examples from MitoEM-R and MitoEM-H datasets.   
 \end{itemize}

\section{Method}
\subsection{Baseline Framework}
We base our approach on the recent Res-UNET~\cite{rsunet}, which utilizes  encoder-decoder structure of 3D UNET~\cite{3dunet} with skip-connections between encoder and decoder. Here, 3D input patch of mitochondria volume ($32\times320\times320$)
 is taken from the entire volume of ($400\times4096\times4096$). The input volume is denoised using an interpolation network adapted for medical images~\cite{huang2020learning}. The denoised volume is then processed utilizing an encoder-decoder structure containing residual anisotropic convolution blocks (ACB). The ACB contains three layers of 3D convolutions with kernels {($1\times3\times3$), ($3\times3\times3$), ($3\times3\times3$)} having skip connections between first and third layers. The decoder outputs  semantic mask and instance boundary, which are then post-processed using connected component labelling to generate final instance masks. We refer to~\cite{rsunet} for more details.

\noindent \textbf{Limitations:} As discussed above, the recent Res-UNET approach utilizes 3D convolutions to handle the volumetric input data. However, 3D convolutions are designed to encode short-range spatio-temporal feature information and struggle to model global contextual dependencies that extend beyond the designated receptive field. In contrast, the self-attention mechanism within the vision transformers possesses the capabilities to effectively encode both local and global long-range dependencies by directly performing a comparison of feature activations at all the space-time locations. In this way, self-attention mechanism goes much beyond the receptive field of the conventional convolutional filters. While self-attention has been shown to be beneficial when combined with convolutional layers for different medical imaging tasks, to the best of our knowledge, no previous attempt to design spatio-temporal self-attention as an exclusive building block for the problem of 3D mitochondria instance segmentation exists in literature. Next, we present our approach that effectively utilizes an efficient spatio-temporal attention mechanism for 3D mitochondria instance segmentation.

\begin{figure}[t!]
\includegraphics[width=\textwidth]{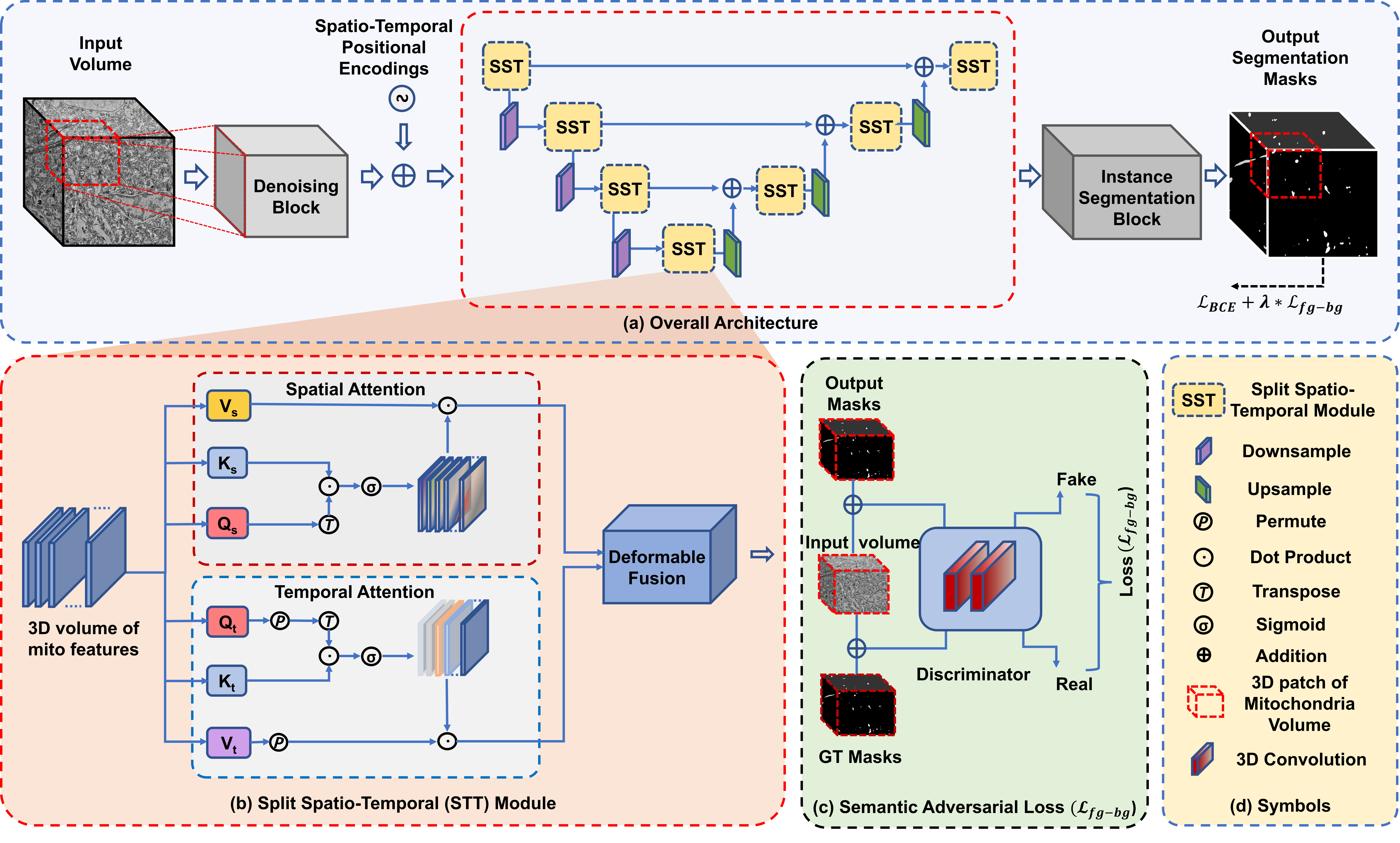} \vspace{-0.7cm}
\caption{\textbf{(a)} Overall architecture of our  STT-UNET framework for 3D mitochondria instance segmentation. A 3D volume patch of mitochondria is first pre-processed using the interpolation network. The resulting reconstructed volume is then fed to 
our split spatio-temporal attention based encoder-decoder to generate the semantic-level mitochondria segmentation masks. The focus of our design is the introduction of split spatio-temporal attention (SST) module within the encoder-decoder. \textbf{(b)} The SST module first computes spatial and temporal attentions independently, which are later combined through a deformable convolution. Consequently, the semantic masks from the decoder are then input to the instance segmentation module to generate the final instance masks. The entire framework is trained using the standard BCE loss ($L_{BCE}$) and our semantic foreground-background (FG-BG) adversarial loss ($L_{fg-bg}$). \textbf{(c)} The $L_{fg-bg}$ loss improves the instance-level features, thereby aiding in the better separability of the region of mitochondria instances from the cluttered background. 
} \label{overall_architecture} 
\vspace{-0.6cm}
\end{figure}

\subsection{Spatio-Temporal Transformer Res-UNET (STT-UNET)}
Fig.~\ref{overall_architecture}(a) presents the overall architecture of the proposed hybrid transformers-CNN based 3D mitochondria instance segmentation approach, named STT-UNET. It comprises a denoising module, transformer based encoder-decoder with split spatio-temporal attention and an instance segmentation block. The denoising module alleviates the segmentation faults caused by anomalies in the EM images, as in the baseline. The denoising is performed by convolving the current frame with two adjacent frames using predicted kernels, thereby generating the resultant frame by adding the convolution outputs. The resulting denoised output is then processed by our transformer based encoder-decoder with split spatio-temporal attention to generate the semantic masks. Consequently, these semantic masks are post-processed by an instance segmentation module using a connected component labelling scheme, thereby generating the final instance-level segmentation output prediction. To further enhance the semantic segmentation quality with cluttered background we introduced semantic adversarial loss which leads to improved semantic segmentation in noisy background. 
 
\noindent \textbf{Split Spatio-Temporal Attention based Encoder-Decoder:}
Our STT-UNET framework comprises four encoder and three decoder layers. Within each layer, we introduce a split spatio-temporal attention-based (SST) module, Fig.~\ref{overall_architecture}(b), that strives to capture long-range dependencies within the cubic volume of human and rat samples. Instead of the memory expensive joint spatio-temporal representation, our SST module splits the attention computation into a spatial and a temporal parallel stream. The spatial attention refines the instance level features from input features along the spatial dimensions, whereas the temporal attention effectively learns the inter-dependencies between the input volume. The resulting spatial and temporal attention representations are combined through a deformable convolution, thereby generating spatio-temporal features. 
As shown in Fig~\ref{overall_architecture} (b), the normalized 3D input volume of denoised features $X$ of size $(T\times H\times W\times C)$ where $T$ is volume size, $(H\times W)$ is spatial dimension of volume and $C$ is number of channels. The spatial and temporal attention blocks project $X$ through linear layer to generate $Q_{s}$, $K_{s}$, $V_{s}$ and $Q_{t}$, $K_{t}$, $V_{t}$. In temporal attention $Q_{t}$, $K_{t}$, $V_{t}$ is permuted to generate $Q_{tp}$, $K_{tp}$, $V_{tp}$ for temporal dot product.  The spatial and temporal attention is defined as,

\begin{equation}
X_s = softmax(\frac{Q_{s}K_{s}^T}{\sqrt{d_k}})V_{s}
\end{equation}

\begin{equation}
X_t = softmax(\frac{Q_{tp}K_{tp}^T}{\sqrt{d_k}})V_{tp}
\end{equation}

Where, $X_s$ is spatial attention map, $X_t$ is temporal attention map and $d_k$ is dimension of $Q_{s}$ and $K_{s}$.
To fuse spatial and temporal attention maps, $X_s$ and $X_t$, we employ deformable convolution. The deformable convolution generates offsets according to temporal attention map $X_t$ and by using these offsets the spatial attention map $X_s$ is aligned. The deformable fusion is given as, 

\begin{equation}
X = \int_{c=1}^{C} \sum_{k_n\in R} W(k_n) \cdot X_s(k_0 + k_n + \Delta K_n) 
\end{equation}

Where, $C$ is no of channels, $X$ is spatially aligned attention map with respect to $X_t$. $W$ is the weight matrix of kernels, $X_s$ is spatial attention map, $k_0$ is starting position of kernel,  $k_n$ is enumerating along all the positions in kernel size of $R$ and $\Delta K_n$ is the offset sampled from temporal attention map $X_t$. We empirically observe that fusing spatial and temporal features through a deformable convolution, instead of concatenation through a conv. layer or addition, leads to better performance. The resulting spatio-temporal features of decoder are then input to instance segmentation block to generate final instance masks, as in baseline.

\noindent \textbf{Semantic FG-BG Adversarial Loss:}
As discussed earlier, a common challenge in mitochondria instance segmentation is to accurately delineate the region of mitochondria instances from the cluttered background. To address this, we introduce a semantic foreground-background (FG-BG) adversarial loss during the training to enhance the FG-BG separability. 
Here, we introduce the auxiliary discriminator network $D$ with two layers of 3D convolutions with stride $2$ during the training as shown in Fig.~\ref{overall_architecture}(c). The discriminator takes the input volume $I$ along with the corresponding mask as an input. Here, the mask $M$ is obtained either from the ground truth or predictions, such that all mitochondria instances within a frame are marked as foreground. While the discriminator $D$ attempts to distinguish between ground truth and predicted
 masks ($M_{gt}$ and $M_{pred}$, respectively), the model $\Psi$ learns to output semantic mask such that the predicted masks $M_{pred}$ are close to ground truth $M_{gt}$. Let $\mathbf{F}_{gt} = \mathrm{CONCAT}(\mathbf{I},\mathbf{M}_{gt})$ and $\mathbf{F}_{pr} = \mathrm{CONCAT}(\mathbf{I},\mathbf{M}_{pred})$ denote the real and fake input, respectively, to the discriminator $D$. Similar to \cite{pix2pix}, the adversarial loss is then given by,

\begin{equation}
\small
L_{fg-bg} = \min_{\Psi} \max_{D} \Psi[\log D(F_{gt})] + \Psi[\log (1 - D(F_{pr}))] + \lambda_1 \Psi[D(F_{gt})-D(F_{pr})]
\end{equation}

Consequently, the overall loss for training is: 
$L = L_{BCE} + \lambda \cdot L_{fg-bg}$, Where, $L_{BCE}$ is BCE loss, $\lambda=0.5$ and $L_{fg-bg}$ is semantic adversarial loss. 

\section{Experiments}

\noindent\textbf{Dataset:} We evaluate our approach on three datasets: MitoEM-R \cite{mitoEM_Dataset}, MitoEM-H \cite{mitoEM_Dataset} and Lucchi \cite{lucchi}. The MitoEM \cite{mitoEM_Dataset} is a dense mitochondria instance segmentation dataset from ISBI 2021 challenge. The dataset consists of 2 EM image volumes (30 $\mu m ^3$) of resolution of $8\times8\times30$ nm, from rat tissues (MitoEM-R) and human tissue (MitoEM-H) samples, respectively. Each volume has 1000 grayscale images of resolution $(4096\times4096)$ of mitochondria, out of which train set has 400, validation set contains 100 and test set has 500 images. Lucchi \cite{lucchi} is a sparse mitochondria semantic segmentation dataset with training and test volume size of $165\times1024\times768$.  

\noindent\textbf{Implementation Details:} We implement our approach using Pytorch1.9 \cite{pytorch} (rcom env) and models are trained using 2 AMD MI250X GPUs.  
During training of MitoEM, for the fair comparison, we adopt same data augmentation technique from \cite{mitoEM_Dataset}. The 3D patch of size $(32\times320\times320)$ is input to the model and trained using batch size of 2. The model is optimized by Adam optimizer with learning rate of $1e^{-4}$. Unlike baseline~\cite{rsunet}, we do not follow multi-scale training and perform single stage training for 200k iterations. For Lucchi, we follow training details of~\cite{rsunet,mitoEM_Dataset} for semantic segmentation. 
For fair comparison with previous works, we use the same evaluation metrics as in the literature for both datasets. We use 3D AP-75 metric~\cite{mitoEM_Dataset} for MitoEM-R and MitoEM-H datasets. 
For Lucchi, we use jaccard-index coefficient (Jaccard) and dice similarity coefficient (DSC).

\begin{table}[t!]
\parbox{.49\textwidth}{
\caption{State-of-the-art comparison in terms of AP on Mit-EM-R and MitoEM-H validation sets. Best results are in bold.} \vspace{-0.2cm}
\label{tab:sota_mitodataset}
\resizebox{\linewidth}{!}{
\begin{tabular}{l|c|c}
\toprule
\rowcolor{mygray} 
\cellcolor{mygray} Methods & MitoEM-R & MitoEM-H \\ \midrule
Wei~\cite{mitoEM_Dataset}  & 0.521 & 0.605 \\
Nightingale~\cite{nightingale2021automatic} & 0.715 & 0.625 \\
Li ~\cite{li2021contrastive} & 0.890 & 0.787 \\
Chen~\cite{rsunet} & 0.917 & 0.82 \\ \midrule
\rowcolor{orange!6} \textbf{ STT-UNET (Ours)} &\textbf{0.958} &\textbf{0.849}  \\ \bottomrule  
\end{tabular}
}
}
\hfill
\parbox{.49\textwidth}{
\caption{State-of-the-art comparison in terms of Jaccard and DSC on Lucchi test set. Best results are in bold.} \vspace{-0.2cm}
\label{tab:sota_lucchidataset}
\resizebox{0.94\linewidth}{!}{
\begin{tabular}{l|c|c}
\toprule
\rowcolor{mygray} 
\cellcolor{mygray} Methods & Jaccard & DSC \\ \midrule
Yuan~\cite{yuan2020net} & 0.865 & 0.927 \\
Casser~\cite{casser2020fast} & 0.890 & 0.942 \\
Res-UNET-R~\cite{rsunet} & 0.895 & 0.945 \\
Res-UNET-R + MRDA ~\cite{chenmicaai2022} & 0.897 & 0.946 \\
\midrule
\rowcolor{orange!6}  \textbf{ STT-UNET (Ours)} & \textbf{0.913} & \textbf{0.962}  \\ \bottomrule
\end{tabular}
}
}
\end{table}

\subsection{Results}
\begin{figure}[t!]
\includegraphics[width=\textwidth]{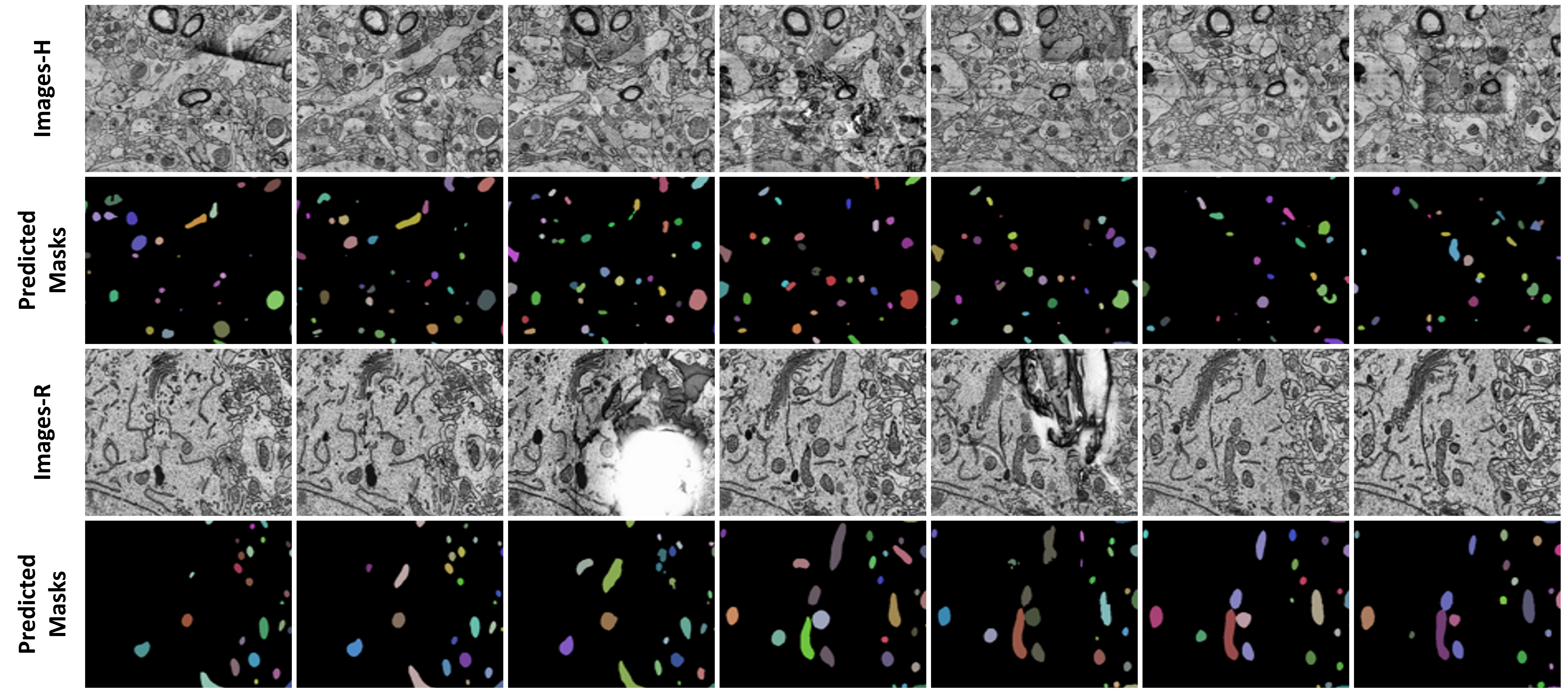} \vspace{-0.7cm}
\caption{Qualitative 3D instance segmentation results of our STT-UNET on the example input regions from MitoEM-H and MitoEM-R val sets. Our STT-UNET achieves promising results on these input examples containing noise. } \label{qual_figure}  
\vspace{-0.2cm}
\end{figure}

\noindent \textbf{State-of-the-Art Comparison:} 
Tab.~\ref{tab:sota_mitodataset} shows the comparison on MitoEm-R and MitoEM-H validation sets. Our STT-UNET achieves state-of-the-art performance on both sets. Compared to the recent~\cite{rsunet}, our STT-UNET achieves an absolute gains of 4.1\% and 2.9\% on MitoEM-R and MitoEM-H sets, respectively. Note that~\cite{rsunet} employs two decoders for MitoEM-H. In contrast, we utilize only a single decoder for both MitoEM-H and MitoEM-R sets, while still achieving improved segmentation performance. Fig~\ref{qual_figure} presents the segmentation predictions of our approach on example input regions from the validation set. Our approach achieves promising segmentation results despite the noise in the input samples. Tab.~\ref{tab:sota_lucchidataset} presents the comparison on Lucchi test set. Our STT-UNET sets a new state-of-the-art on this dataset in terms of both Jaccard and DSC. 

\begin{table}[t!]
\setlength{\tabcolsep}{1.1pt}
\begin{minipage}{.33\linewidth}
\centering
\caption{ Baseline performance comparison. } \vspace{-0.2cm}
\label{tab:ablation_baseline_comp}
\resizebox{0.85\linewidth}{!}
{
\begin{tabular}{l|c|c}
\toprule
\rowcolor{mygray} 
\cellcolor{mygray} Methods & MitoEM-R & MitoEM-H \\ \midrule
Baseline  & 0.921 & 0.823 \\
+ SST & 0.948 & 0.839 \\
\rowcolor{orange!6} + $L_{fg-bg}$ & \textbf{0.958} & \textbf{0.849} \\
 \bottomrule
\end{tabular}
}
\end{minipage}\hfill
\begin{minipage}{.33\linewidth}
\centering

\caption{Ablation study on the impact of feature fusion. } \vspace{-0.2cm}
\label{tab:ablation_featur_fusion}

\resizebox{\linewidth}{!}
{
\begin{tabular}{l|c|c}
\toprule
\rowcolor{mygray} 
\cellcolor{mygray} Feature Fusion & MitoEM-R & MitoEM-H \\ \midrule

addition  & 0.950 & 0.841 \\
concat & 0.952 & 0.842 \\ 
\rowcolor{orange!6} \textbf{def-conv} & \textbf{0.958} & \textbf{0.849}  \\ \bottomrule  
\end{tabular}
}
\end{minipage}\hfill
\begin{minipage}{.33\linewidth}
\centering

\caption{ Ablation study on the impact of design choice. } \vspace{-0.2cm}
\label{tab:ablation_design_choice}

\resizebox{1.08\linewidth}{!}
{
\begin{tabular}{l|c|c}
\toprule
\rowcolor{mygray} 
\cellcolor{mygray} Deisgn choice & MitoEM-R & MitoEM-H \\ \midrule

spatial-temporal  & 0.922 & 0.817 \\
temporal-spatial & 0.937 & 0.832\\ 
\rowcolor{orange!6} \textbf{spatial|temporal} & \textbf{0.958} & \textbf{0.849} \\ \bottomrule  
\end{tabular}
}
\end{minipage} 
\end{table}

\noindent \textbf{Ablation study:} 
Tab.~\ref{tab:ablation_baseline_comp} shows a baseline comparison when progressively integrating our contributions: SST module and semantic foreground-background adversarial loss. The introduction of SST module improves performance from 0.921 to 0.941 with a gain of 2.7\%. The performance is further improved by 1\%, when introducing our semantic foreground-background adversarial loss. Our final approach achieves absolute gains of 3.7\% and 2.6\% over the baseline on MitoEM-R and MitoEM-H, respectively.
Tab.~\ref{tab:ablation_featur_fusion} shows ablation study with feature fusion strategies in our SST module: addition, concat and deformable-conv. The best results are obtained with deformable-conv on both datasets.
For encoding spatial and temporal information, we analyze two design choices with SST module: cascaded and split, as shown in Tab.~\ref{tab:ablation_design_choice}. The best results are obtained using our split design choice (row 3) with spatial and temporal information encoded in parallel and later combined.
We also evaluate with different input volumes: 4,8,16,32. We observe best results are obtained when using 32 input volume. 


\section{Conclusion}
We propose a hybrid CNN-transformers based encoder-decoder approach for 3D mitochorndia instance segmentation. We introduce a split spatio-temporal attention (SST) module to capture long-range dependencies within the cubic volume of human and rat mitochondria samples. The SST module computes spatial and temporal attention in parallel, which are later fused. Further, we introduce a semantic adversarial loss for better delineation of mitochondria instances from background. Experiments on three datasets demonstrate the effectiveness of our approach, leading to state-of-the-art segmentation performance. 

\bibliographystyle{splncs04}
\bibliography{bibfile}

\end{document}